\documentstyle[sprocl,twoside,epsf]{article}

\def\citebk#1{\hspace{0.9mm}\raisebox{-1.85mm}[0mm][0mm]
  {\Large\cite{#1}}\hspace{-0.1mm}}

\def\citebkcap#1{\hspace{0.8mm}\raisebox{-1.5mm}[0mm][0mm]
  {\large\cite{#1}}\hspace{-0.2mm}}

\begin{document}

\thispagestyle{plain}
\setcounter{page}{1}

\title{QUANTUM TUNNELING \\[2mm]
In Memory of M. Marinov}

\author{M. SHIFMAN}

\address{Theoretical Physics Institute, University of Minnesota,
Minneapolis,
MN 55455}

\maketitle

\vspace*{0.4cm}

\abstracts{This article is a slightly
expanded version of  the talk I delivered
at the Special Plenary Session
of the 46-th Annual Meeting of the Israel Physical Society
(Technion, Haifa, May 11, 2000) dedicated to Misha Marinov. 
In the first part I briefly discuss quantum tunneling, 
a topic which Misha cherished and to
which  he was repeatedly returning through his career.
My task was to show that Misha's work had been deeply woven 
in the fabrique of  today's theory.
The second part is an attempt to highlight one of many facets of Misha's
human portrait. In the 1980's, being a {\em refusenik} in Moscow,
he volunteered to teach physics under unusual circumstances.
I  present recollections of people who were involved in this activity.}

\vspace*{0.5cm}

\tableofcontents

\newpage
\section{Introduction}

I am honored to give this talk at the special session of the Israel 
Physical Society. A brief summary of  
Marinov's life-long accomplishments in mathematical and high-energy
physics was given by Professor Lipkin.
 My task was to choose one theme, from several
which Misha considered to be central in his career,
to show how deeply it is intertwined in the fabrique of  today's
theory. I was asked to prepare a nontechnical presentation
that would be understandable, at least, in part, to nonexperts.   

 The center of gravity of 
Marinov's research interests  rotated around advanced aspects of
quantum  mechanics, quasiclassical quantization and functional
integration  methods. For this event I have chosen the
 topic of  quantum tunneling, which Misha cherished and to
which  he was repeatedly returning through his career, the last time in
1997.\,\cite{1} The reader interested in a more technical discussion
of Misha's results on quantum tunneling is referred
to Segev's and Gurvitz's articles in this Volume.\cite{S,G}

Apart from the direct involvement in
the tunneling-related projects, this subject carries an imprint of
other Marinov's contributions --
from dynamics of the Grassmannian variables\,\cite{2}
to the monopole studies  in gauge theories.\cite{3}
I will explain this shortly. The second part of my talk is nonscientific.
It is a sketch, an attempt to present one of many facets of Misha's
human portrait. In the 1980's, being a {\em refusenik} in Moscow,
he volunteered to teach physics under unusual circumstances.
I conducted interviews with people who were involved in this activity.
\vspace*{-2mm}

\section{What is quantum tunneling?}

To demonstrate the essence of the phenomenon,
let us consider a simple example -- the $\alpha$ decay of heavy nuclei.
For definiteness, one can keep in mind the decay
$$
U_{92}^{238} \to Th_{90}^{234} +\alpha\, .
$$
The energy of the emitted $\alpha$ particle in this decay
is 4.7 MeV.
The emitted $\alpha$ particle and the daughter $Th_{90}^{234}$
nucleus experience an interaction. At large distances this is the Coulomb 
repulsion
\begin{equation}
V_{\rm C} = \frac{Z_1Z_2e^2}{r} = \frac{2\times 90 \times
e^2}{r}= 
\frac{260}{r/\mbox{fermi}
}\, \mbox{ MeV}\,.
\end{equation}
The Coulomb potential is equal to 4.7 MeV at $r=55$ fermi.
At such distances the nuclear force (which is attractive) is negligible
since its range approximately coincides with the nucleus radius,
\begin{equation}
R_n \approx 1.2 A^{1/3} = 1.2\times(234)^{1/3}\approx 7.4\,\mbox{fermi}\,.
\end{equation}
A sketch of the corresponding potential is 
shown in Fig.\,1, where $V_{\rm C}^*\approx35\,$MeV.
At $r\!>\!R_n$ the potential is purely Coulombic, while
at $r\!<\!R_n$ it is essentially determined by the nuclear interaction
and is flat in a rough approximation. The energy of the
$\alpha$ particle is denoted
by $E_1$.
\begin{figure}[h]   
\epsfxsize=8cm
\centerline{\epsfbox{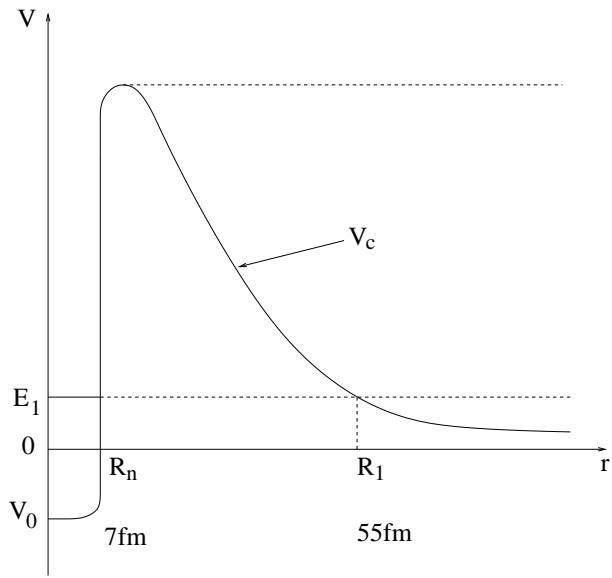}}
 \caption{A sketch of the potential energy for $\alpha$ particle
in the $\alpha$ decay of $U_{92}^{238}$.}
\end{figure}
At the initial moment of time the $\alpha$ particle is confined in the
well at $r\!<\!R_n$. According to the laws of the classical
mechanics it will stay there forever -- the potential barrier at
$R_n\!<\!r\!<\!R_1$
prevents the classical particle from the leakage 
in the domain to the right of $R_1$. Thus, classically $U_{92}^{238}$
is stable with respect to the $\alpha$ decay.

In actuality, the laws of quantum mechanics do allow the
$\alpha$ particle to leak under the barrier. This is called ``Quantum 
Tunneling." The wave function of the $\alpha$ particle
spreads under the barrier even if initially
it was confined in the well to the left of $R_n$. If the barrier is high,
as is the case in the problem at hand, the tunneling probability
is exponentially small, and it can be calculated quasiclassically.
The tunneling amplitude is proportional
to\,\footnote{~Here and below I use units in which $\hbar = 1$,
a standard convention in particle physics.} 
\begin{equation}
A_{\rm tunneling} \propto \,\exp \left\{ -\!\int_{R_n}^{R_1}\!\!\!\!
\sqrt{2m (V-E_1) }\,
dr\right\}
.
\label{simple}
\end{equation}
 The integral in the exponent runs over the classically forbidden domain.
The decay probability is given by $|A|^2$; it can be readily evaluated 
using
the simple expression (\ref{simple}) which nicely explains the fact that
the probability is extremely small, of order $10^{-37}$ in the problem at 
hand, and is extremely sensitive to the $\alpha$ particle energy.
For all known $\alpha$-particle emitters, the value of $E_1$ varies
from about 2 to 8 MeV. Thus, the value of the $\alpha$ particle energy 
varies only by a factor of 4, whereas the
range of  lifetimes (inverse probabilities)
is from about $10^{11}$  years down to about $10^{-6}$ second, a factor
of $10^{24}$. 

The quantum tunneling as an explanation of the heavy nuclei decays  
was suggested by George Gamow\,\cite{4} in 1928 and, somewhat later,
(but  independently) by Gurney and Condon.\cite{5} 

\begin{figure}[h]   
\epsfxsize=6cm
\centerline{\epsfbox{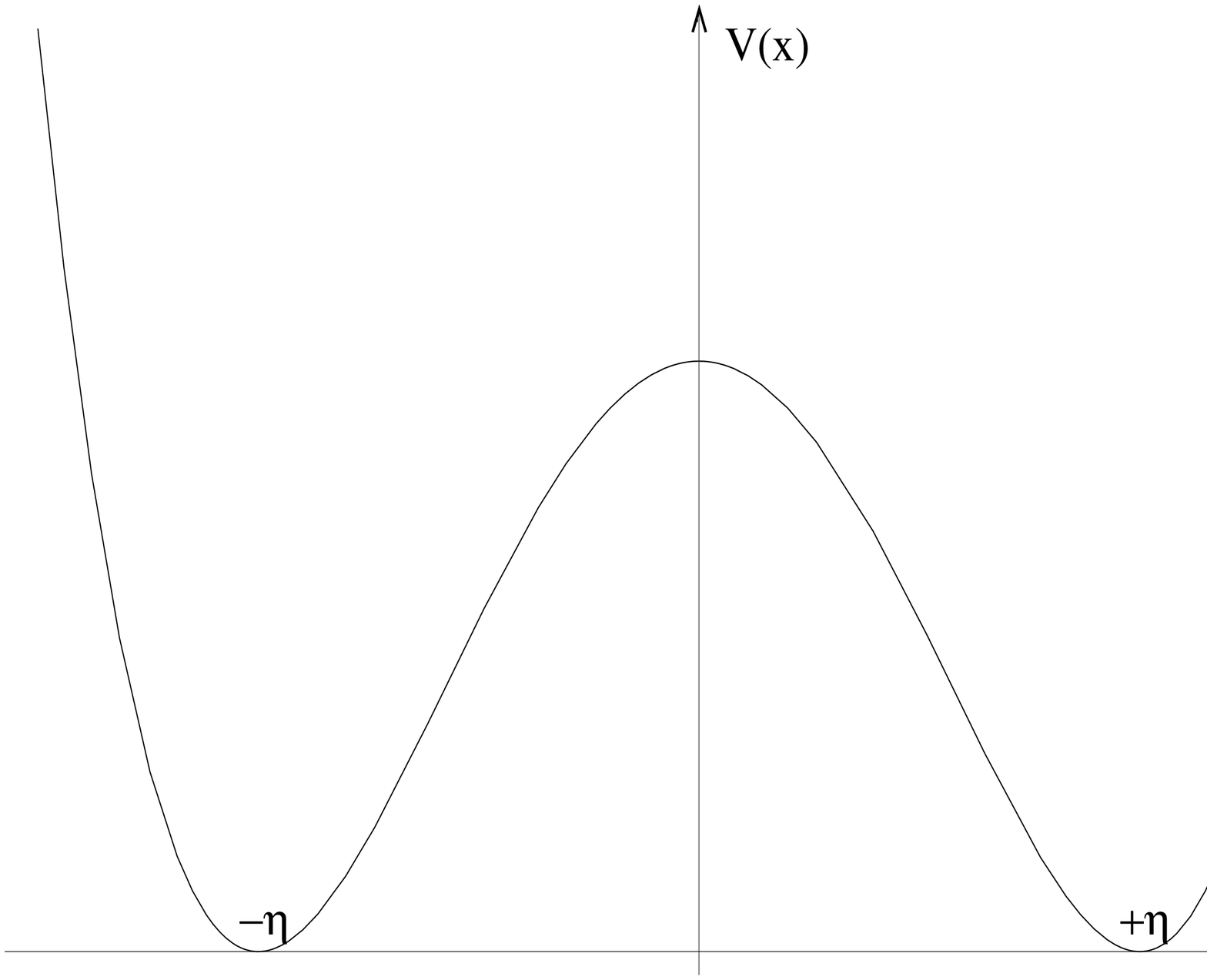}}
 \caption{The  double-well potential.}
\end{figure}

Quantum tunneling is at work in a much wider range
of  phenomena than the heavy 
nuclei decays just discussed.
I chose this example for illustrative purposes since in these
processes the consequences of the tunneling can be 
readily explained to nonexperts. The very same tunneling
determines a global structure of the ground state in a wide variety of 
quantal and field-theoretical problems.  For instance, in the text-book 
problem  of the double-well potential (see Fig.\,2), there is no instability.
The tunneling from the left to the right well and {\em vice versa}
manifests itself in that the ground state is unique
(rather than doubly degenerate as would be the case in the
classical theory). It is symmetric with respect to 
$x\!\to\!-\,x$,  and there is an exponentially small energy splitting
between the low-lying  $P$-even and $P$-odd states. This splitting is
determined by the very same expression (\ref{simple}) where one may
put
$E_1\!=\!0$, and the integral runs from $-\eta$ to $\eta$. 
It can be represented in an identical form as
\begin{equation}
A_{\rm tunneling} \propto \exp\,(-S_0)
\,,
\label{qsimple}
\end{equation}
where $S_0$ is the action on the classical trajectory $x (\tau )$
connecting the points $-\eta$ to $\eta$ in the distant past and distant 
future in the imaginary time. In the real time there are no classical
trajectories connecting $-\,\eta$ and $\eta$, the domain between these 
two points is classically forbidden. The particle tunneling  can be 
described as particle's motion   in the imaginary time,
$t \!\to\!-\,i \tau$. Passing to the imaginary time, $t\!\to\!-\,i \tau$,  we 
effectively change the sign of the potential, see Fig.\,3. The double-well 
potential becomes two-hump. 
It is pretty obvious that in the two-hump potential there is a classical
trajectory 
\begin{equation}
m\,\ddot{x} = -\,\frac{dV}{dx }\, , \qquad V(x) = -\lambda
\left (x^2\!\!-\eta^2\right)^2
\end{equation}
interpolating between $-\,\eta$ and $\eta$. The minimal action is
\begin{equation}
S_0 = \frac{m^2\omega^3}{12\lambda}\,.
\end{equation}

\begin{figure}[h]  
\epsfxsize=6cm
\centerline{\epsfbox{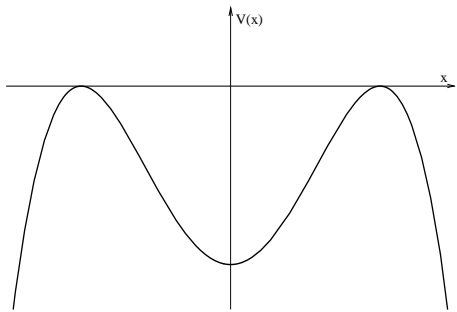}}
 \caption{The potential relevant to
the imaginary time classical motion in the double-well problem.}
\end{figure}

This strategy  is  general in the problems of this type. The
tunneling  amplitude is determined (in the quasiclassical approximation)
by the minimal action attained on the least-action trajectory
interpolating through 
 the classically forbidden domain in the imaginary time.
In one-dimensional problems determination of such trajectories and 
calculation of the corresponding minimal action
is a relatively trivial task. The task becomes exceedingly more 
complicated 
in multi-dimensional 
quantal problems and in field theory.  Much of Marinov's time and 
effort  was  invested in the theory of the multi-dimensional tunneling. 

\section{Multi-dimensional quantal problems}

Now I would like to demonstrate, in some graphic examples, that
70 years after its inception  the
topic is alive and continues to attract theorists' attention.
The formulation of the problem
 to be discussed below is inspired by supersymmetry,
although the corresponding  dynamical systems 
need not be supersymmetric. Supersymmetry, by the way, was also
very close to Marinov's heart. The early work he did with
Berezin\,\cite{2} on the classical description  of particles with spin
in terms of the Grassmannian variables carries elements
inherent to supersymmetry. Moreover, the last paper on which Misha
worked being already terminally ill, was a historic essay on
supersymmetry.\cite{6}

In this section I will consider a class of quantal multi-dimensional
problems. Assume that we have $n$ degrees of freedom,
$X^1, X^2, ..., X^n$, described by the Hamiltonian
\begin{equation}
H = \sum_{\ell =1}^n\, \frac{1}{2}\, \big(\dot{X^\ell}\big)^2\!+\,
V\!\big(X^\ell\big)\,,
\end{equation}
where the potential $V$ can be represented as
\begin{equation}
V = \frac{1}{2}\,\sum_{\ell =1}^n \left( \frac{\partial\, W\!
({X})}{\partial X ^\ell}
\right)^2,
\label{six}
\end{equation}
 $W$ is a function of all variables ${X} ^\ell$, which I will
call ``superpotential", bearing in mind its origin in supersymmetric
theories. The mass $m$ is put $m\!=\!1$. This is done for simplicity,
and is by no means necessary. In
the given context the term superpotential is nothing but a convenient
name for the function $W$. 

Certainly, the potential (\ref{six}) is very special. It has an interesting
feature that at the extremal points of the superpotential,
\begin{equation}
 \frac{\partial\, W\! ({X})}{\partial X ^\ell}
=0\,,\qquad \ell = 1,2,...,n\,,
\label{seven}
\end{equation}
the potential $V$ vanishes. Since the potential (\ref{six})
is positive-definite these are the classical ground states of the system.
Let us assume that  Eq.\,(\ref{seven}) has more than one solution,
and denote these solutions by
$\xi_{a}\! \equiv \!\{\xi^\ell \}_{a} $ where the subscript  labels distinct solutions,
\begin{equation}
 \frac{\partial\,W\!({X}^\ell)}{\partial X^\ell}
=0\,, \quad\mbox{at} \quad X^\ell =\left( \xi^\ell\right)_{a}\,,\quad a=1,2,...
\label{eight}
\end{equation}
For simplicity I will assume that there are only two solutions,
$\xi_1$ and $\xi_2$. This assumption does not lead to the loss of
generality, it just makes the notation simpler. The barrier separating
the points $\xi_1$ and $\xi_2$ is assumed to be high (see Fig.\,4),
so that the tunneling problem is solvable in the quasiclassical
approximation.

\begin{figure}[h]  
\epsfxsize=8cm
\centerline{\epsfbox{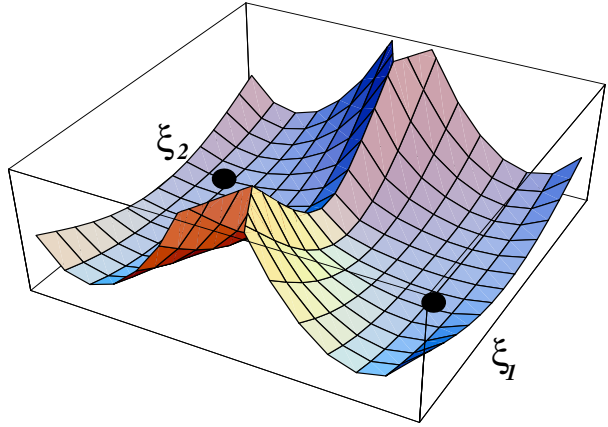}}
 \caption{A profile of potential energy with two degenerate minima
at $\xi_1$ and $\xi_2$. The barrier separating these two points
is assumed to be high, so that tunneling can be described quasiclassically.}
\end{figure}

The tunneling amplitude from the first to the second classical minimum
of the potential $V$ is given by Eq.\,(\ref{qsimple}),
but finding the tunneling trajectory 
and the  minimal action $S_0$ in case
of  a significant number of variables, $n\gg 1$, 
 does not seem to be a trivial task, especially if the
potential $V$ is contrived. 

In fact, I will show momentarily that $S_0$ can be obtained\,\cite{7}
without solving the classical equations of motion for the trajectory,
\begin{equation}
S_0 = \left|W(\xi_1) -W(\xi_2)\right| .
\label{nine}
\end{equation}

The classical equations in the imaginary time
take the form
\begin{equation}
\ddot{X}^\ell = \frac{\partial\, V\!({X})}{\partial X^\ell}\, .
\label{ten}
\end{equation}
This is a second order differential equation.
Consider, instead, the following first order differential equation
\begin{equation}
{\dot X}^{\ell} = -\frac{\partial\,W\!({X})}{\partial X^\ell}\, .
\label{eleven}
\end{equation}
If we manage to find a solution of this latter equation,
it will be automatically the solution of Eq.\,(\ref{ten}). Indeed,
differentiating both sides of Eq.\,(\ref{eleven})
over $\tau$ one gets
\begin{equation}
\ddot{X}^\ell = -\sum_j\frac{\partial^2 W }{\partial
X^\ell \, \partial X^j} {\dot X}^j =
\sum_j\frac{\partial^2 W }{\partial
X^\ell \, \partial X^j} \,
 \frac{\partial W }{\partial X^j} =  \frac{\partial\,V\!(X)}{\partial
X^\ell}\, .
\label{twelve}
\end{equation}
(The inverse is not true).

Equation (\ref{eleven}) has a very graphic physical interpretation --
it describes the motion of a very viscous fluid (like honey)
on the multi-dimensional profile given by
$W(X_\ell )$.  The motion must originate in one extremal point of $W$
(which, obviously, must be a local maximum or a saddle point) and end up 
in the second extremal point (which must be a local minimum or
a saddle point).  The left-hand side is the velocity,
there is no inertia; the right-hand side is the ``force." The velocity is
proportional to the force. That's how honey flows.

 Although it may be
difficult to find the flow trajectory analytically for contrived profiles, 
a rich mechanical  intuition all of us have
regarding this type of motion tells us immediately
whether the solution exists or there is an obstruction,
i.e. the honey flow from the summit is diverted away from the
minimum, into an abyss. Once we see that the solution exists, 
we can get $S_0$ ``for free",
\begin{equation}
S_0 = \frac{1}{2}\!\int\!\! d\tau
\bigg(\!\sum_\ell \dot{X^\ell} \dot{X^\ell} + V(x)\!
\bigg)\! \equiv  \frac{1}{2}\!\int\!\! d\tau\!
\left\{\!
\bigg(\!\sum_\ell\!\dot{X^\ell}\!+  \frac{\partial W }{\partial X^\ell}
\bigg)^2\!\!\!-2\,\frac{dW}{d\tau}\!
\right\}.
\label{thirteen}
\end{equation}
The expression in the parentheses on the right-hand side 
vanishes on
the solution, while the second term, the full time derivative,
yields Eq.\,(\ref{nine}). 

It may happen that the flow equations (\ref{eleven})
do not have a solution with the proper asymptotics --
honey does not want to flow from the $a$-th
to the $b$-th extremum of $W$. The second order equations 
(\ref{twelve}) always  do
have a solution, however. From (\ref{thirteen}) we see that in this case
$
S_0>\! |W(\xi_b) -W(\xi_a)| \,,\,
$
and 
\begin{equation}
\exp\big(-\left|W(\xi_b) -W(\xi_a)\right|\big) 
\label{fourteen}
\end{equation}
presents an upper bound on the tunneling amplitude.

The set-up described above (with the first
order equations replacing the second order equations of motion)
is called
the Bogomol'nyi-Prasad-Sommer\-field (BPS) saturation.\cite{Bo,PS}
BPS saturated objects are in the focus of today's research in
supersymmetric theories -- from supersymmetric gluodynamics to
string and brane theory, where the BPS saturation
 acquires a deep
symmetry foundation. The BPS saturated objects preserve 1/2 or 1/4 of
the original supersymmetry. The set-up was introduced by
Bogomol'nyi in the pre-supersymmetry era.
In 1975 Zhenya Bogomol'nyi was a Ph.D. student at ITEP, like myself.
Immediately after the discovery of the 't Hooft-Polyakov 
monopoles he started calculating their masses. The first
calculation was numerical, a joint project with Misha Marinov.\cite{3}
In this work Bogomol'nyi and Marinov  observed that the limit of the
vanishing Higgs self-coupling was special; in this limit the
monopole mass tended to $4\pi M_W\!/g^2$, with the coefficient
exactly equal to 1.  This observation served as a hint and  an initial
impetus for the subsequent work,\cite{Bo}
where the limit of the vanishing
Higgs self-coupling was worked out analytically,
and the monopole mass was reduced to a surface term, the magnetic
charge.

\vspace{0.2cm}

\section{Dynamics on manifolds}

The above example describes dynamics generated by
the potential $V\!(X)$ related to the superpotential $W\!(X)$
through Eq.\,(\ref{six}). The space is flat, $X^\ell\!\in\! R_n$.
In fact, one can readily generalize the problem to include
the motion on nontrivial manifolds ${\cal T}$, endowed with the metric tensor
$g_{ij} (X^\ell ) $,
\vspace{0.1cm}
\begin{equation}
L = \frac{1}{2}\,  g_{ij}  \dot{X}^i \dot{X}^j
- \frac{1}{2}\, g^{ij}\frac{\partial W}{\partial X^i }\,\frac{\partial W}{\partial X^j }\,,
\label{crvdl}
\end{equation}
where the coordinates $X^\ell$ parametrize  ${\cal T}$,
and  $W\!(X^\ell)$ is a function on ${\cal T}$. As is evident from
Eq.\,(\ref{crvdl}), the formula for the potential 
that replaces Eq.\,(\ref{six}) is
\begin{equation}
V(X) = \frac{1}{2}\, g^{ij} \, \frac{\partial W}{\partial X^i}\frac{\partial W}{\partial X^j}\,.
\label{crvdp}
\end{equation}
I have already mentioned in the beginning of my talk that  
quantization on group manifolds (a particular class of nontrivial
${\cal T}$'s) was Misha's perennial topic. Together with Terentev, he wrote
a famous article on this subject.\cite{novem}

The BPS equations in this case take the form
\begin{equation}
\frac{d X^i}{d\tau} = - g^{ij}\frac{\partial W}{\partial X^j}\,,
\label{crvd}
\end{equation}
with the boundary conditions that the trajectory starts
at one minimum of the potential $V$ and ends at another
(i.e. interpolates between two distinct extremal points of $W$).
The solution of this equation interpolating between $\xi_a$ and $\xi_b$,
if it exists, determines the minimal action and the tunneling
amplitude (in the exponential approximation).
As previously, the minimal action can be inferred
without the knowledge of the explicit solution, and is given by 
the same formula,
see Eq.\,(\ref{nine}). A remarkable feature is its
{\em metric-independence}.

The existence/nonexistence of the solutions of Eqs.\,(\ref{crvd})
depends on the global (topological) characteristics of the manifold ${\cal T}$
and the superpotential $W$. The corresponding theory -- it 
goes under the name of the Morse theory -- is a well-developed 
and quite beautiful branch of mathematics
(see e.g. Refs.\,\protect\citebk{novem2} and \protect\citebk{novem3}).
Unfortunately, I have no time to dwell
on it, even briefly. The best I can do is to offer a simple illustration.

Let us assume that ${\cal T}$ is a two-dimensional sphere $S_2$,
and the superpotential is such that it has two isolated extremal points,
one at the north pole and another at the south (Fig.\,5).

\begin{figure}[h]  
\epsfxsize=4cm
\centerline{\epsfbox{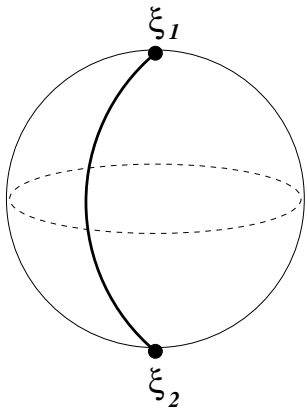}}
 \caption{Tunneling trajectories for the particle on the 
sphere. The thick line is the flow line of the
honey droplet from one extremal point of $W$
to another.
}
\end{figure}

If one extremal point is the maximum of $W$, another one has to be the minimum.
Alternatively, they can be saddle points. In both cases it is obvious that
a droplet of honey, being placed in one extremal point, will find 
its way to another --
the solution of the BPS equations always exists. In the first case
(maximum at the north pole and  minimum at the south), there
exists a continuous family of solutions 
-- the flow lines of  the droplet can 
leave the pole along any meridian.

 Now, one can start deforming
${\cal T}$ and $W$. For instance, one can squeeze the
sphere of Fig.\,5 transforming it into the surface of, say, a melon or a lemon, or 
another
surface.
 As far as the existence of the BPS solution
is concerned, nothing will change for small continuous deformations of ${\cal T}$ and
$W$. The particular form of the solution will change, of course, but not 
the fact of its existence. 

The solution connecting the
north and south poles may disappear only if the deformations 
will be so strong that either  topology of ${\cal T}$  changes or
$W$   develops new extremal points which might  trap the droplet
on its way from one pole to another. If there are several extremal points,
${\cal T}$ splits into several domains -- they are called the Morse cells.
Each cell $M_{ab}$ presents a domain swept by the family of trajectories
connecting $a$ and $b$,
$$
{\cal T} =\sum_{\rm pairs} M_{ab}\,.
$$

\section{Beyond physics}

I would like to conclude my talk on a personal note.
I first saw Misha Marinov when I came to ITEP around 1969 to pass my 
entrance examinations. The  examination session was chaired by
  Professor V.B.~Berestetskii. I do not remember exactly what 
topic was under discussion. I remember, however, that when a certain  
subtle point was raised, Berestetskii interrupted the session and said
that  at this point 
it would be appropriate to consult Okun. We went to his office, which 
was across the hall, and Marinov was there. Okun introduced us.

Since then we would occasionally bump into each other in the corridors of 
ITEP, or in the seminar hall. In
1979  Misha became a {\em refusenik}, a nonperson 
according to the Orwellian nomenclature. I remember that this news --
that Marinov had applied for emigration to Israel -- came when many of 
us were at the Odessa School of Physics  organized by the Landau 
Institute.  I heard Okun telling somebody 
 that Marinov had left ITEP.
In those days such a step actually meant ``disappearance beyond the 
horizon".  In 1986, in the beginning  of {\em perestroika}, when
Marinovs finally got their exit visas to Israel, I came to their place to say
{\em farewell}, since I was absolutely sure that I would never ever see 
any of 
them again (as were many other friends and colleagues who gathered in 
tiny Marinov's apartment on Bolotnikovskaya street that night). 

Well, life can be much richer than our imagination.  It was not Marinov
who was forced to give up his  dream to live and die on the land of his
ancestors, but, rather, the  antihuman system which kept him hostage
for years and to which he   resisted, collapsed. 

Many people in this audience knew Misha from the human side, as 
a person on whom one
could always rely on a rainy day,  always open to one's needs and ready to help. 
I'd like to say a few words on that, dwelling on one of Misha's activities
which, regretfully,  remains virtually unknown to public.
In Marinov's obituary in {\em Physics Today} (the September 2000 issue)
it is mentioned that in the early 1980's  ``...  he
volunteered to teach physics at the unofficial university organized by {\em
refuseniks} to educate Jewish students who were barred by authorities from
regular universities."  

In fact, Misha volunteered to teach  at the
so-called ``Jewish People's University" in Moscow which 
was organized in the late 1970's 
by Bella   Abramovna Subbotovskaya. At this time discrimination against
Jews in the admission policies of the Soviet universities 
reached its peak. Sure enough, it was not the first peak, but it was  
especially strong. In fact, virtually all Departments of  Mathematics
in Soviet Union completely 
closed their doors to Jewish students.\footnote{~Statistical data 
illustrating this fact in the most clear-cut manner were
presented in  {\em samizdat} essays, Refs.\,\protect\citebkcap{A} and 
\protect\citebkcap{B}, 
and in Freiman's book.\cite{C}
I am grateful to Natasha Zanegina, Gabriel Superfin, and Gennady Kuzovkin
for providing me with these materials.}
 I do not know why, but (this is a
well-known fact) the Soviet mathematical community, which gave to
the world some of the most outstanding mathematicians, was also
pathologically anti-Semitic. Such great mathematicians as Pontryagin and
Vinogradov, who had enormous administrative powers in their hands,
were ferocious anti-Semites. The tactics used for cutting off
Jewish students were very simple. At the entrance examination
special  groups of ``undesirable applicants" were organized.\footnote{~The
working definition of ``Jewishness" was close to that of Nazis; 
 having at least one Jewish parent of even grandparent  would almost
certainly warrant  one's placement in the category of undesirables.} They
were then offered killer  problems which, as a rule, were at the level of
international mathematical competitions; sometimes, 
they were flawed, or
even completely wrong.\footnote{~An extremely illuminating analysis of
such killer problems which had been collected in Ref.\,\protect\citebkcap{D}
was published by Ilan Vardi.\cite{E} Another
brief collection of the killer problems can be found in Appendix B in
Ref.\,\protect\citebkcap{C}.}

Bella Subbotovskaya's idea was to launch something like unofficial
extension classes, 
where children unfairly  barred from the
official universities could get food for their the hungry minds  from the
hands of the first-class mathematicians and physicists.
It was supposed that the classes would take place on a regular basis
through the entire school year, that they would be open to everybody (no
registration or anything of this kind was required) and that the spectrum
of courses offered would be broad and deep enough to provide a serious
educational background in the exact sciences.

 Andrei Zelevinsky
(now at the North-Eastern University, Boston) recollects:
\begin{quote}
``Bella  Abramovna was  the main organizer [...].
I didn't know her very well: she was much older than me [...]. It was 
she who
brought Fuchs and me to teach there. I was truly impressed with her courage
and  quiet
determination to run the whole thing. All the organizational work, 
from
finding the places for our regular
meetings to preparing sandwiches
for  participants
was done by her and two other activists: Valery Senderov and [...] 
Boris Kanevsky. Both as I understand, were active dissidents at the time. 
I think they made a deliberate effort to separate mathematics from
politics,  in order to protect us,  professional research
mathematicians."
\end{quote}
Another professor of this ``university'', Dmitry Fuchs (now at UC, Davis),
writes: 
\begin{quote}
``We taught there major mathematical disciplines corresponding
to  the
first two years  of the Mekh-Mat 
 curriculum: mathematical analysis, linear algebra and geometry, abstract
algebra and so on. I taught there since 1980 through 1982, until
Subbotovskaya's  sudden death
in September or October of 1982 put an end to this enterprise.
Bella Abramovna was tragically killed in a hit-and-run 
accident
which was universally believed to be the act of KGB. This has  never been
officially confirmed though. At this time we did not use any particular name
for our courses. In spirit,  this was more a continuous 
seminar rather  than a
university. Among other teachers, I knew Andrei Zelevinsky, Boris Feigin,
and Alexei  Sossinski [...]. The number of students varied between 60 and 20.
The  place of our meetings was not permanent: we met at an elementary
school where  Bella Abramovna  worked as a teacher, at the Gubkin
Institute for Petrol  and Gas,
 at the Chemistry and Humanities Buildings of the Moscow State 
University, and so on. Anywhere, where we could get permission to occupy a
large enough room\ldots .   I personally asked the
 Deputy Dean of the  Chemistry
Department  for such a permission, and it was granted.
All  instructors prepared notes which
 were photocopied  and distributed among the students. An article on our
``school" was published by a Russian-Israeli newspaper some time ago.
It was interesting but excessively emotional. Among 
other things, the authors had a tendency to exaggerate the Jewish nature of
our ``university." It is certainly 
true that a substantial part of both, students
and  teachers,
were ethnic Jews. This was the result of the well-known 
policy of the 
Mechmat admission committee,  and the Soviet State at large, rather than
the deliberate aim  of the organizers. After all, we taught only
exact sciences. No plans were made to teach Jewish
culture, history, or language."
\end{quote}
Needless to say that all teachers were enthusiasts, and they received
no reward other than the wonderful feeling that 
what they were doing was a good deed,
a {\em mitzva}.  Misha Marinov taught  topics in physics.

Although the goals of this  ``university" were purely educational,
 the very fact of its existence
was considered by the authorities as a political act of resistance. 
Any association with
it was dangerous. The tragic fate of Bella Subbotovskaya
 remains uninvestigated so far. Perhaps, in the future her life will
become a subject of a historical treatise, or a novel, who knows?
Another activist who was mentioned above by Andrei Zelevinsky, Valery
Senderov, was sentenced to 7 years in jail with the subsequent 5-year exile to
Siberia on charges of anti-Soviet agitation and propaganda  in 1983.
One of the main pieces of incriminating evidence supporting the  charges
was his and Kanevsky's essay {\em Intellectual Genocide.}\cite{B} If 
one reads this 
essay now it is hard to understand how a person could be sent to prison
for describing  mathematical  problems suggested 
to particular applicants at the entrance
examinations at the Department of Mathematics of the Moscow University.
Well, they say that everything goes  in the struggle of
classes. Why not mathematics?

But let me return to Misha and his 
generous contribution to this endeavor.
A former student of Misha, 
Boris Shapiro  recollects (his article\,\cite{BSh} is published in  this Volume):
\begin{quote}
``Soon after the beginning of the lectures on mathematics, Bella
Abramovna announced that for all who desire to come, there would be a
series of lectures by a famous physicist on quantum mechanics and field 
theory.[...]
The very next Thursday, I think,  in the tiny two-room apartment of Bella
Abramovna on Nametkina Street, eight or nine people met, mainly
second-year students [...].  There
appeared a tall, sinewy gray-haired man, with a slight lisp.  
The man was
Misha Marinov, the name by which 
he introduced himself.  By his
appearance, Misha was close to forty-five years old; he had already
been a {\em refusenik} for some time, and had seen a great deal.  He had,
until he applied for his exit visas, worked in ITEP  and was personally 
acquainted
with famous people.  [...]  For me, he
represented that heroic epoch when Jews were still accepted into
the realm of physics and mathematics, where they uncovered so
much.  My impression was reinforced by occasional Misha's stories of
L.D.~Landau, whose biography the younger generation was also eagerly reading.
He spoke of his cooperation with F.A.~Berezin, and it was then that
I first heard of ``supermanifold'' [...]
On the other hand, once (during a tea
break), he told us about
a team of construction workers, {\em shabash\-niks-refuseniks}, to 
which he belonged,
consisting entirely of Ph.D.\  holders and professors.
 Those were dark times, and
it was uncertain when and to where they would let him go.  It was for
this reason that Misha was very careful not to compromise the entire
system of the ``Jewish University'' and other teachers, more or less
well off and at that time far from any desire to immigrate.  As they
say, ``some are no longer with us, and the others are far away\ldots''. 
\end{quote}

The following quotation from Shapiro's essay gives an idea of Misha's choice of
physics topics for teaching. It speaks also of his attitude,
thoughtful and demanding. This was his usual attitude.

\begin{quote}
 ``Misha's lectures were not simple.  An understanding
of mechanics, functional analysis, differential equations,
differential geometry and the like was necessary.  Despite this, I
remember the ease and clarity with which he explained the fundamental
ideas behind the difficult make-up of quantum mechanics.  He pointed
out the similarities and complexities of classical mechanics such that
even I had the impression that I understood everything.  I remember
the interesting discussions on harmonic and anharmonic oscillators
  and the difficulties in the quantization of the latter.
He even managed to discuss with the second-year students
``supermathematics'' on the basis of Grassmann algebra.  [...]  He tried 
with all his strength to
bring us to the level of current research, though this was not very
realistic, meeting once a week with second-year students.  In this
way two intensive years went by, until in the spring of 1982 we were
all stunned by the death of Bella Abramovna.  Everyone went around
very crushed, and the ``Jewish University'' soon fell apart."
\end{quote}

\section*{Acknowledgements}
\addcontentsline{toc}{section}{\numberline{}Acknowledgements}

I am grateful to G.~Eilam and J.~Feinberg  for inviting me to deliver this
talk at the Special Session of the Israel Physical Society. Special thanks go
to A.~Losev for thorough discussions of the Morse theory.
The work was 
supported in part by DOE grant DE-FG02-94ER408.

\section*{References}
\addcontentsline{toc}{section}{\numberline{}References}

\end{document}